\documentclass[aps,11pt,onecolumn,preprintnumbers,amsmath,amssymb]{revtex4}

\usepackage[english]{babel}
\usepackage{graphicx}
\usepackage{color}

\begin{document}

\newcommand{\unit}[1]{\:\mathrm{#1}}            
\newcommand{\To}{\mathrm{T_0}}
\newcommand{\Tp}{\mathrm{T_+}}
\newcommand{\Tm}{\mathrm{T_-}}
\newcommand{\EST}{E_{\mathrm{ST}}}
\newcommand{\Rp}{\mathrm{R_{+}}}
\newcommand{\Rm}{\mathrm{R_{-}}}
\newcommand{\Rpp}{\mathrm{R_{++}}}
\newcommand{\Rmm}{\mathrm{R_{--}}}
\newcommand{\ddensity}[2]{\rho_{#1\,#2,#1\,#2}} 
\newcommand{\ket}[1]{\left| #1 \right>} 
\newcommand{\bra}[1]{\left< #1 \right|} 

\title{Generation of heralded entanglement between distant hole spins}
\author{Aymeric Delteil$^{1*}$}
\author{Zhe Sun$^{1*}$}
\author{Wei-bo Gao$^{1,2*}$}
\author{Emre Togan$^{1}$}
\author{Stefan Faelt$^{1}$}
\author{Ata\c c Imamo\u glu$^{1\dagger}$}
\affiliation{$^1$Institute of Quantum Electronics, ETH Zurich, 8093 Zurich, Switzerland \\
$^2$Div. of Physics and Applied Physics, Nanyang Tech. Univ., Singapore 637371 \\
$^*$These authors contributed equally to this work. \\
$^\dagger$Corresponding author. E-mail: imamoglu@phys.ethz.ch}


\maketitle \textbf{Quantum entanglement emerges naturally in
interacting quantum systems  and plays a central role in quantum
information processing~\cite{Nielsen, Briegel98, Lukin96, Kimble08}.
Remarkably, it is possible to generate entanglement even in the
absence of direct interactions: provided that which path information
is erased, weak spin-state dependent light scattering can be used to
project two distant spins onto a maximally entangled state upon
detection of a single photon \cite{Cabrillo98}. Even though this
approach is necessarily probabilistic, successful generation of
entanglement is heralded by the photon detection event. Here, we
demonstrate heralded quantum
entanglement~\cite{Monroe07,Weinfurter12,Blatt13,Hanson13} of two
quantum dot heavy-hole spins separated by 5 meters using
single-photon interference. Thanks to the long coherence time of
hole spins and the efficient spin-photon interface provided by
self-assembled quantum dots~\cite{Gao12,deGreeve12,Shaibley13}
embedded in leaky microcavity structures, we generate 2300 entangled
spin pairs per second, which represents an improvement approaching
three orders of magnitude as compared to prior
experiments~\cite{Monroe15}. Delayed two-photon interference scheme
we developed allows for efficient verification of quantum
correlations. Our results lay the groundwork for the realization of
quantum networks in semiconductor nanostructures. Combined with
schemes for transferring quantum information to a long-lived memory
qubit~\cite{Meyer15}, fast entanglement generation we demonstrate
could also impact quantum repeater architectures.}

In contrast to prior experiments demonstrating electron spin photon
entanglement \cite{Gao12,deGreeve12,Shaibley13}, our experiments are
based on heavy-hole pseudo-spins in self-assembled quantum dots (QD)
that have been shown to exhibit long coherence times
\cite{Warburton,Yamamoto, Greilich, Carter}. Figure~1a depicts our
experimental set-up incorporating two QDs separated by 5 meters that
are resonantly driven by weak $3.2$~ns long pulses from a
Ti:Sapphire laser, termed the entanglement laser. Additional diode
laser pulses ensure that each QD is optically charged with a single
excess heavy-hole and that the hole pseudo-spin is prepared in the
requisite state. The QDs are embedded in distributed Bragg reflector
(DBR) structures~\cite{Senellart13} that allow for efficient
collection of the generated resonance fluorescence.

Figure~1b shows the relevant energy-level diagram as well as the
allowed optical transitions for single-hole charged QDs when an
external magnetic field ($B_x$) is applied perpendicular to the
growth direction (Voigt geometry) \cite{Xu07,Atature06}. The initial
states of the optical transitions in the single-hole charged regime
are metastable states identified by the orientation of the hole
pseudo-spin, with $| \Uparrow \rangle$ ($|\Downarrow \rangle$)
denoting hole angular momentum projection. Presence of $B_x \neq 0$
yields a finite splitting of the pseudo-spin states due to
heavy-light hole mixing~\cite{Bayer02}. Spontaneous emission of a V
(H) polarized photon at frequency $\omega_{blue}$ ($\omega_{diag1}$)
from the trion state $| T_b\rangle$ at rate $\Gamma/2$ brings the QD
back into the $| \Downarrow \rangle$ ($|\Uparrow \rangle$) state.
Due to these selection rules, addressing any of the four allowed
transitions with a single laser will efficiently transfer the spin
population into the opposite ground state within a few optical
cycles. Since the intensity of entanglement laser is chosen to be
well below saturation, the ensuing optical transitions lead to
either V-polarized Rayleigh scattering or H-polarized Raman
scattering.

The light propagation time from the first beam splitter (BS1) to
both dots, as well as from the dots to the second beam splitter
(BS2) are rendered nearly identical, such that the photons scattered by the
two dots during a single entanglement laser pulse recombine at the
same time on the second beam splitter. When both QDs are initially
prepared in the $| \Downarrow \rangle$ state, the simultaneous weak
excitation of the blue transitions will lead to either a spontaneous
Raman or Rayleigh scattering event with a probability $\varepsilon^2
\ll 1$, leaving the system in the state
\begin{equation} |\Psi\rangle_{12} = \frac{1}{\sqrt{2}}[ | \Downarrow , 0 \rangle
+ \varepsilon e^{-i\theta_1}(| \Uparrow , 1_{H} \rangle + |
\Downarrow , 1_{V} \rangle)]_{QD1}\otimes[| \Downarrow , 0 \rangle +
\varepsilon e^{-i\theta_2}(| \Uparrow , 1_H \rangle + | \Downarrow ,
1_V \rangle)]_{QD2}.
\end{equation}

To ensure that a click in one of the single-photon detectors stem
form detection of a H-polarized Raman scattered photon at frequency
$\omega_{diag1}$, we placed transmission gratings and Fabry-P\'erot
filters in front of the detectors (Fig.~1a). In this case, detection
of a single (Raman) photon projects the composite system
wave-function onto the maximally entangled state
\begin{equation}
\label{entangledstate} |\Psi\rangle_{12} = \frac{1}{\sqrt{2}}[ |\Uparrow, \Downarrow
\rangle +e^{-i\theta} |  \Downarrow, \Uparrow  \rangle]
\end{equation}
in the limit where two photon scattering probability $\varepsilon^4$
is vanishingly small. Provided that the Zeeman splitting in the two
QDs are rendered identical, the relative phase $\theta = \theta_2 -
\theta_1$ is time-independent and is primarily determined by the
optical path length difference between the two arms from BS1 to BS2
(Fig.~1a).

The entanglement generation  scheme we use relies crucially on the
indistinguishability of the photons emitted by two remote QDs such
that "which-path" information is not available in the single-photon
interferometer depicted in Fig.~1a. After locating a pair of QDs (QD1 and QD2) with similar transition energies, we ensure that the optical
transition frequencies $\omega_{blue}$ and $\omega_{diag1}$ are
identical for QD1 and QD2; this is achieved by tuning both electric
and magnetic fields applied on the two QDs separately. The
indistinguishability of the Raman scattered photons is characterized
by a Hong-Ou-Mandel (HOM) experiment~\cite{HOM}  under pulsed excitation of the
blue transition where we monitor the coincidence events at the two
single photon detectors. Figure~1c shows that two-photon
interference results in a strong decrease of the two-fold
coincidence rate associated with the central peak when the two input
photons have parallel polarizations. The associated interference
visibility deduced from this measurement is $91\pm 6$~\%.

For the protocol we implement, it is essential that the QD spins
remain coherent during the time it takes for the heralding process
to be completed. In our scheme this time is determined predominantly
by the $21.7$~ns propagation time from the QDs to the single-photon
detectors. To demonstrate that the hole pseudo-spin retains its
coherence on this timescale, we implemented a quantum optical
measurement technique. It is well known in quantum optics that while
first-order coherence properties of Rayleigh scattering follows that
of the excitation laser \cite{Loudon,Atature}, the coherence of
spin-flip Raman scattering is determined both by the laser and the
spin coherence~\cite{Fernandez09}. The latter is a consequence of
the fact that the quantum field $E^{(+)}(t)$ generated  in Raman
scattering is linearly proportional to the spin raising operator
$\sigma_{\Uparrow \Downarrow} (t-R/c)$, where $R$ denotes the
distance between the QD and the detector. Therefore, hole spin
coherence ($T_2^*$) time can be determined by measuring the
coherence time of Raman scattered photons, provided that the
excitation laser has a much longer coherence time.

To perform this experiment, we use the set-up depicted in Fig.~2a,
where the emitted photons at a desired wavelength are filtered  and
then sent into a stabilized Mach-Zehnder interferometer whose path
length difference is set to 22~ns. We apply the pulse sequence
depicted in Fig.~2b: the spin is first prepared in the $ |
\Downarrow \rangle $ state by spin pumping using the red (vertical)
transition. We then apply two weak pulses on the blue transition,
separated by the same time as the path length difference, such that
the light scattering amplitudes during the two pulses can interfere
at the second beam splitter. The detector is then gated so that it
measures only in this interference time window and the photon
detection events are recorded as a function of the phase difference
between the two arms. As a reference, we measure an interference
visibility of 83~\% for the excitation laser, limited by the
precision with which we stabilize the optical path length
difference. For Raman photons we obtain an interference visibility
of 38.9~\% for QD1 and 29.5~\% for QD2, which is more than half the
visibility measured for Rayleigh scattering (66.7~\% for QD1 and
54.8~\% for QD2). The visibility of Rayleigh photons is limited by a
contribution from incoherent light scattering, whereas Raman
coherence is in addition reduced by the decay of spin coherence.
This result demonstrates that both spins retain a sizeable degree of
coherence during a time-window of 22~ns and that the reduced
visibility is mainly due to the contribution from incoherent light
scattering.

The fringes in the aforementioned Mach-Zehnder interferometer could also be observed by
varying the phase of the hole pseudo-spin. The latter can be
adjusted using a V-polarized off-resonant laser field that induces
different phases on the two spin states due to different magnitude
of the ac-Stark effect. We carried out this experiment on QD1 by
applying a laser that is red-detuned by $\sim20$~GHz from the red
and $\sim50$~GHz from the blue vertical transition (bottom right
diagram of Fig.~2d). The difference in the ac-Stark shift
experienced by the two transitions allows the state $ | \Downarrow
\rangle$ to accumulate a phase $\varphi = \Omega^2 \tau \delta / 4
\Delta (\Delta + \delta)$, relative to $ | \Uparrow
\rangle$.  Here, $\Omega$ is the Rabi frequency of
the laser, $\tau$ the pulse duration, $\Delta$ the detuning from the
red transition and $\delta = \omega_{blue}-\omega_{red}$ the energy
difference between the two transitions. In order to characterize the
effect of spin-phase rotation, we repeat the interferometric
measurement while keeping the optical path-length difference constant and
applying a 4~ns-long detuned laser pulse in between the two weak
excitation pulses (Fig.~2d). By varying the laser power from 0 to
$\sim2~\mu$W, we change the relative phase of the two spin states
and thus the relative phase of the Raman scattering amplitude before
and after the pulse that induces the spin-state dependent ac-Stark shift. The oscillations in the count rate as a
function of the laser power (Fig.~2e, black dots) unequivocally
demonstrate single pseudo-spin rotation about the $z$ axis of the
Bloch sphere. The red curve in Fig.~2e is a sinusoidal fit to the
data, showing that no sizable loss of visibility is observed for
spin rotation up to $4 \pi$.

Having demonstrated that hole spin coherence persists for longer
than 22 ns, we address the verification of heralded spin-spin
entanglement. To demonstrate classical correlations between the
distant spins, we carry out local single-spin measurement in the
computational basis, conditioned upon the detection of a Raman
photon during the entanglement pulse. We benefit from the fact that
each spin state can be excited to a corresponding trion state with
the same oscillator strength and the same laser polarization but
using a different resonant laser wavelength. The detection of a
photon during a blue (red) laser pulse thus tells with a high
confidence level that the state of the spin prior to the measurement
pulse was $ |\Downarrow \rangle$ ( $ |\Uparrow \rangle$). In order
to measure the four different spin combinations under the same
experimental conditions, we alternate in a single experiment four
pulses sequences, each  performing one of the four requisite
measurement combinations. The full pulse sequence is described in
Fig.~3a: we first prepare the state $ |\Downarrow,
\Downarrow\rangle$ by spin pumping, then apply the weak entanglement
laser pulse. The detection of a Raman photon during this pulse
heralds successful entanglement generation. We then successively
measure the state of the two dots. The measurement pulses of the two
dots are offset in time, allowing us to extract which-path
information. All the measurements are performed close to saturation
and the detection efficiencies are rendered similar. The duration of
the full sequence is $4\times104$~ns. Figure~3b shows the results of
the 3-fold coincidences detected during 106.5 hours of measurement.
As expected, the odd parity events, where the spins of the two dots
are opposite, are much more likely than the even parity events,
where the two spins are found in the same state. The associated
fidelity is $F_z = 80.6 \pm 6.6$~\%.

To demonstrate quantum correlations between the two distant spins,
we implement a delayed two-photon interference experiment. The key
element of this approach for  verifying quantum correlations is the
possibility to rotate one of the spins along the z-axis after
heralded spin entanglement is generated. Application of a detuned
laser pulse on QD1 as described earlier, results in rotating the
phase of the entangled state by $\alpha (\tau)$ so that the the
entangled state becomes $(| \Uparrow ,\Downarrow \rangle
+e^{-i\theta-i\alpha(\tau)} | \Downarrow , \Uparrow
\rangle)/\sqrt{2}$. Subsequent application of a second weak
(measurement) pulse, that is identical in intensity and duration to
the entanglement pulse, on both QDs simultaneously leads to
\begin{eqnarray}
|\Psi\rangle_{12} &=& \frac{1}{\sqrt{2}}[ |\Uparrow ,\Downarrow , 0 \rangle +
\varepsilon e^{-i\theta_2}| \Uparrow ,\Uparrow, 1_H \rangle +
e^{-i\theta-i\alpha(\tau)}(\varepsilon e^{-i\theta_1}
| \Uparrow ,\Uparrow , 1_H \rangle +
| \Downarrow ,\Uparrow , 0 \rangle )] \\
&=& \frac{\varepsilon}{\sqrt{2}} e^{-i \theta_2} ( 1 +
e^{-i\alpha(\tau)})
 | \Uparrow, \Uparrow, 1_H \rangle + \frac{1}{\sqrt{2}} \left(
 |  \Uparrow,\Downarrow , 0\rangle +
e^{-i \theta-i\alpha(\tau)} | \Downarrow ,\Uparrow , 0 \rangle  \right) .
\label{finalState}
\end{eqnarray}
Therefore, conditioned on an initial Raman photon detection event
that heralded spin-spin entanglement,  the detection of a second
time-delayed Raman photon detection probability scales as
$\varepsilon^2 |1 + e^{-i\alpha(\tau)}|^2$. The expectation value of
Raman photon detection can be shown to be
\begin{equation} \langle E^{(-)} E^{(+)} \rangle \propto 1 + 0.5 \langle \sigma_z^1 + \sigma_z^2 \rangle  - \langle
\sigma_{\Downarrow \Uparrow}^2 \sigma_{\Uparrow \Downarrow}^1 +
 \sigma_{\Downarrow \Uparrow}^1 \sigma_{\Uparrow
\Downarrow}^2 \rangle.
\end{equation}
The peak-to-peak contrast in $\langle E^{(-)} E^{(+)} \rangle$
obtained by varying $\alpha(\tau)$ therefore  gives us the magnitude
of non-local quantum correlations between the two spins.

To verify the presence of quantum correlations using such a delayed
two-photon  interference experiment, we use the pulse sequence
described in Fig.~3c. We once again prepare the spins in the $|
\Downarrow ,\Downarrow \rangle$ state by spin pumping and then apply
the entanglement generation pulse. The phase of the state is then
modified by the detuned laser pulse, whose duration is changed
within the pulse sequence by alternating eight patterns that differ
only by the duration of this particular pulse. Eight evenly
distributed durations are chosen to cover more than one full
revolution. Finally the measurement pulse is simultaneously sent to
both dots. The two-fold coincidences measured for each value of the
pulse length are normalized by the uncorrelated coincidence rate
obtained by measuring two photons emitted in different periods. The
figure~3d presents data obtained during 180~min of measurement. The
obtained ratio exhibits clear oscillations of visibility $29.6 \pm
2.8$~\%. Combining the results depicted in Figs.~3b$\&$3d, we deduce
an overall fidelity of the generated entangled state of $F = 55.2
\pm 3.5$~\%.

The non-local measurement of quantum correlations does not allow us
to determine the relative phase $\theta$ between the $|\Uparrow
\Downarrow\rangle$ and $|\Downarrow \Uparrow\rangle$ contributions.
In the ideal limit of z-basis measurements yielding vanishing
probability for $|\Uparrow \Uparrow\rangle$ and $|\Downarrow
\Downarrow\rangle$ states, the state space of the two qubits is
restricted to a two-dimensional subspace which can be mapped  onto a
Bloch sphere with $|\Uparrow \Downarrow\rangle$ and $|\Downarrow
\Uparrow\rangle$ as the north and the south poles. A perfect
visibility in non-local two-photon interference in return ensures
that the two-spin state lives on the equator of this Bloch sphere:
$|\Psi\rangle_{12} = (| \Downarrow , \Uparrow\rangle +e^{-i\theta} |
\Uparrow , \Downarrow \rangle)/\sqrt{2}$. To fix the value of
$\theta$, we need to use a strong laser to stabilize the optical
path length difference to an integer multiple of its wavelength
immediately prior to the generation of the entangled spin state. We
have implemented such a stabilization scheme  to perform the
Mach-Zehnder interferometry measurements depicted in Fig.~2. We
emphasize that it is straightforward to extend our heralded
entanglement generation experiments by fixing $\theta$ with an
accuracy of $\pm \pi/10$ using this stabilization scheme while
achieving nearly the same entanglement generation rate.

Our work establishes QD heavy-hole spin states as ideal candidates
for quantum information processing tasks requiring a spin-photon
interface: combination of long $T_2^*$ coherence times and high
photon collection efficiency yielding up to 10 million photons per
second on saturated exciton transitions, yields an unprecedented
spin-spin entanglement generation rate of 2300 ebits per second.
This rate  could be further increased by a factor of 10 using cavity
QED~\cite{Senellart13}. Increasing the distance between the
entangled spin pairs could be achieved either by using dynamical
decoupling or by using singlet-triplet qubits of QD molecules that
exhibit longer spin coherence times while simultaneously allowing
for efficient spin measurement \cite{Weiss12}. Our results provide
major motivation for investigating possibilities for coherent
transfer of the hole spin to a local qubit that exhibits a longer
coherence time~\cite{Meyer15} but lacks an efficient interface to
propagating optical photons~\cite{hanson11}. A hybrid system that
uses QD spins for fast entanglement generation and long-lived qubits
for storage could have a major impact on implementation of quantum
repeaters~\cite{Lukin96}. In addition, heralded spin entanglement
could provide a quantum coherent link between nodes of an on-chip
quantum information processor \cite{finley,Lodahl}.

\

{}

\vspace{1 cm}

\newpage
\begin{figure}[t]
   \includegraphics[scale=0.5]{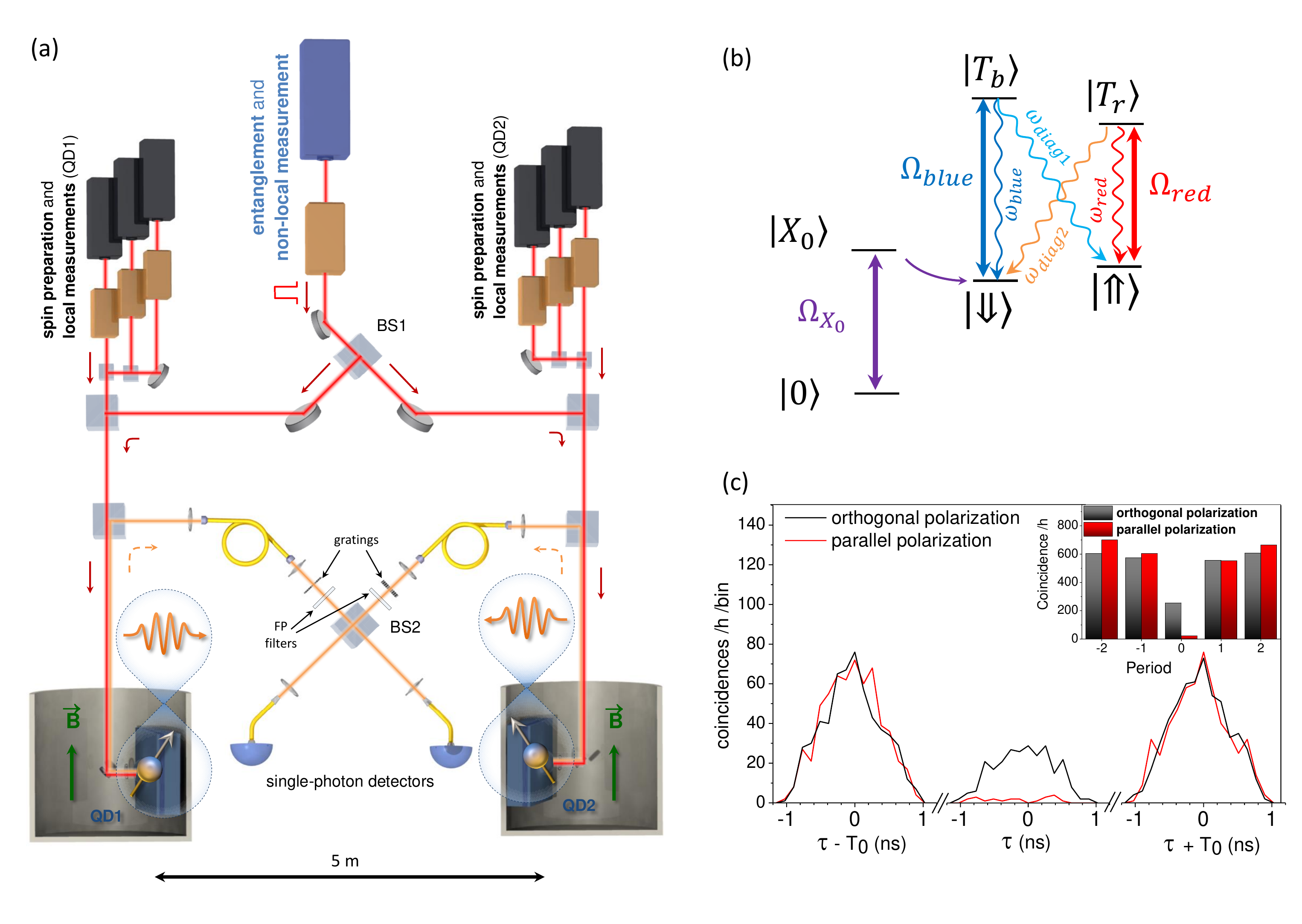}

\textbf{Figure 1. Experimental set-up.}
 \textbf{(A)} Two bath cryostats separated by 5 meters host quantum
dot samples  in Voigt geometry . The quantum dots can be addressed
by diode lasers (in black) for local state preparation and readout,
and by a Ti:Sapphire laser (in blue) for entanglement generation and
non-local measurement. \textbf{(B)} Energy level diagram of a single
quantum dot. Upon excitation of the neutral exciton ($| X^0\rangle$)
state, the electron can tunnel out, leaving behind a single hole.
Application of a finite magnetic field gives rise to spin-dependent
optical selection rules with four allowed transitions of identical
oscillator strength. \textbf{(C)} Characterization of the
indistinguishability of the Raman photons from the two dots with a
Hong-Ou-Mandel experiment: coincidence counts on the two output arms
of BS2 are plotted as a function of the delay between the recorded
photon arrival times under pulsed excitation. $T_0$ is the
repetition period of 52~ns. When the input modes have parallel
polarizations (red curve), the coincidence counts within the time
window (-1~ns, 1~ns) are 11 times smaller than for the case of the
input modes having orthogonal polarization.
\end{figure}

\newpage

\begin{figure}[t]
   \includegraphics[scale=0.6]{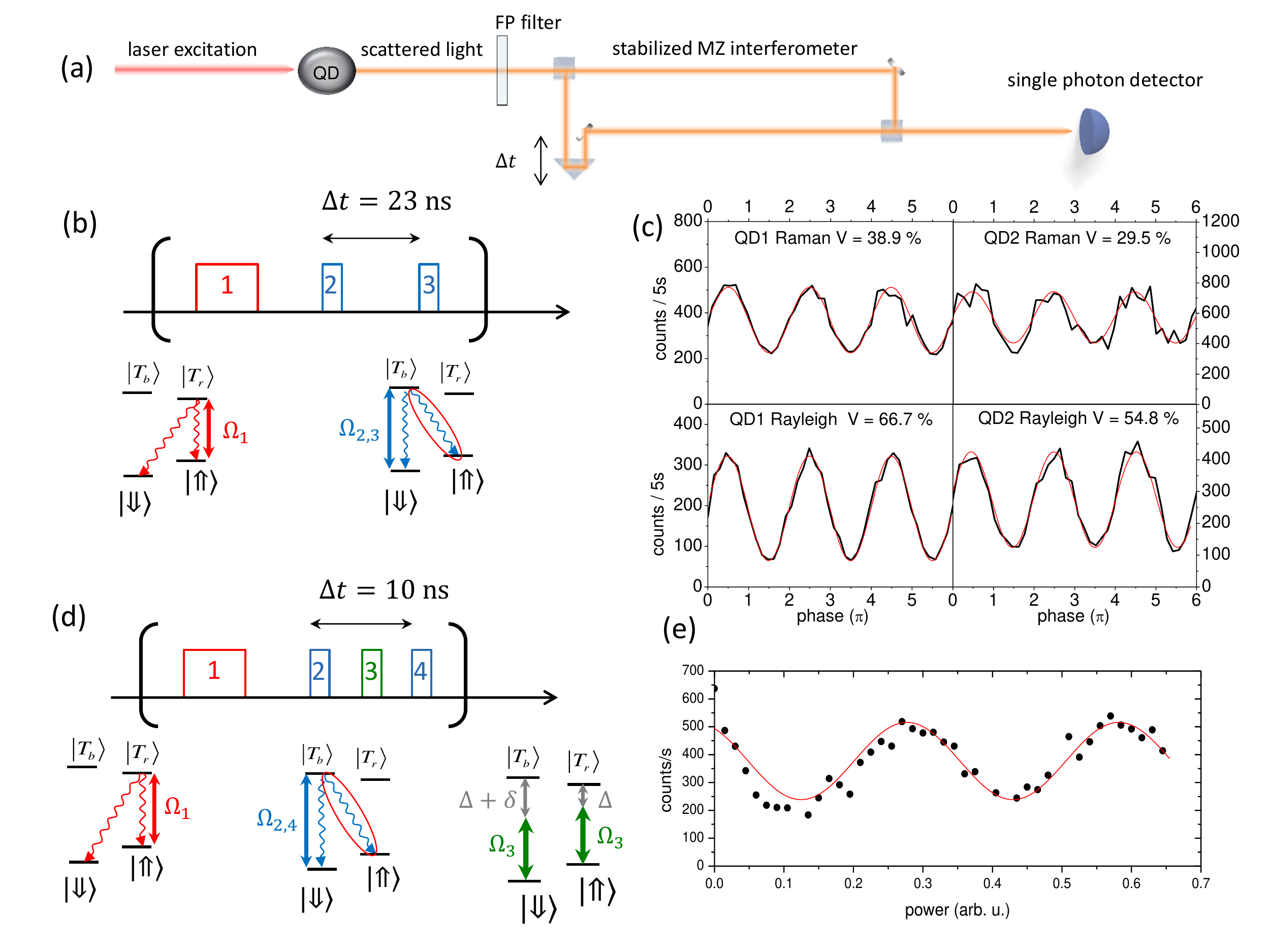}

\textbf{Figure 2. Coherence and rotation of the hole pseudo-spin.}
\textbf{(A)} Sketch of the experimental set-up for the single-photon
interference of Raman scattered photons. The single photons are sent
to an interferometrically stabilized Mach-Zehnder (MZ)
interferometer of path length difference $c \Delta t$. A single
photon detector is placed in one of the output modes of the second
beam-splitter. \textbf{(B)} Pulse sequence used for the first order
coherence measurement, and the relevant energy level diagram. We
first apply a pulse of frequency $\omega_{red}$ to spin pump into
the $| \Downarrow \rangle$ state (pulse~1). We then apply two
successive weak pulses (2 and 3) at frequency $\omega_{blue}$. The
time offset of the two pulses approximately matches the path length
difference of the MZ interferometer. \textbf{(C)} Count rate of the
single photon detector, for QD1 (left column) and QD2 (right
column), when filtering only the Raman (upper row) or the Rayleigh
(lower row) scattered photons, as a function of the phase difference
in the two arms. The associated visibility, obtained from a
sinusoidal fit of the count rate, is indicated in the associated
panel. \textbf{(D)} Pulse sequence used to demonstrate pseudo-spin
rotation about the $z$ axis of the Bloch sphere: the pulse sequence
is identical to the one used for part (B), with an additional
detuned laser pulse of 4~ns (pulse~3) is inserted in between the two
pulses at $\omega_{blue}$ (2 and 4). \textbf{(E)} Black dots: count
rate of the output detector, as a function of the detuned laser
power, demonstrating control of the pseudo-spin phase. Red curve:
fit to the data. \end{figure}

\newpage

\begin{figure}[t]
   \includegraphics[scale=0.6]{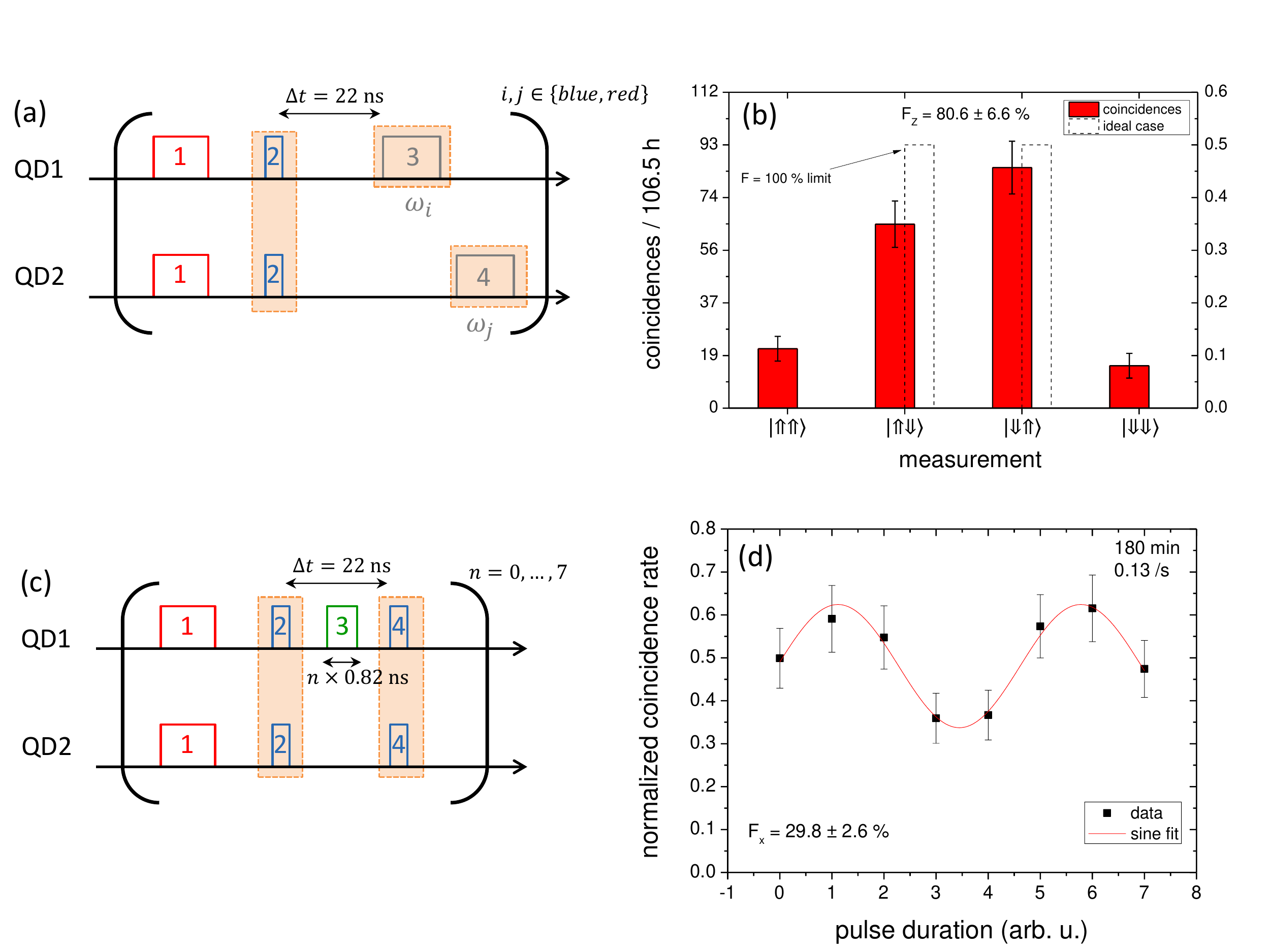}

\textbf{Figure 3. Characterization of the heralded entangled state.}
\textbf{(A)} Pulse sequence used for the  measurement of classical
correlations between the distant spins. After spin pumping into the
$| \Downarrow, \Downarrow \rangle$ state (pulse~1), a weak
entanglement pulse (pulse~2) is sent simultaneously to both quantum
dots (QD1 and QD2). After 22~ns, pulse~3 measures the spin-state
ofQD1 and then pulse~4 measures the spin-state of QD2. The four
measurement combinations are alternated. The total repetition period
of the measurement processes is $4 \times 104$~ns. \textbf{(B)} Red
bars: results of three-fold coincidences between a photon emitted
during the entanglement pulse and a photon in each of the two
measurement pulses (orange shading in Fig.~3a) obtained during a
total measurement time span of 106.5~h. The dashed bars represents
the ideal limit of vanishing even parity spin state detection. The
error bars represent one standard deviation deduced from poissonian
statistics of the raw detection events. The measured fidelity is
$F_z = 80.6 \pm 6.6$~\%. \textbf{(C)} Pulse sequence used to measure
quantum correlations between the distant spins. After spin pumping
into the $| \Downarrow, \Downarrow \rangle$ state, a weak
entanglement pulse (pulse~2) is used to drive both quantum dots. A
detuned laser pulse (pulse~3) modifies the phase of the QD1 hole
spin phase. After 22~ns, a non-local measurement pulse is applied to
both QDs. The pulse sequence is repeated for different values of the
duration of the pulse~3 ranging from 0 to $8\times0.82$~ns,
corresponding to a laser-induced QD1 spin phase rotation ranging
from $0$ to $3 \pi$. \textbf{(D)} Black dots: two-fold coincidence
rate between a photon detected during the entanglement pulse and a
second photon detected during the measurement pulse (orange shading
in Fig.~3c), normalized by the average detection rate between
photons emitted during different periods, as a function of the
pulse-length of pulse~3. The error bars represent one standard
deviation deduced from poissonian statistics of the raw detection
events. The red curve is a sinusoidal fit to the data, yielding a
visibility of $V = 29.8 \pm 2.6$~\%.
\end{figure}
\end{document}